\documentclass{article}
\usepackage[utf8]{inputenc}
\usepackage{natbib}
\usepackage{graphicx}
\usepackage{listings}
\usepackage{amsmath}
\usepackage{color}
\usepackage{courier}
\usepackage{textcomp}
\usepackage{amsfonts, amssymb}
\usepackage{subcaption}
\usepackage{multirow}
\usepackage{booktabs}
\usepackage{hyperref}

\title{Optimizing Block-Sparse Matrix Multiplications on CUDA with TVM}
\author{Zijing Gu\\
\texttt{zijingg@cs.cmu.edu}}
\date{\today}

\begin{document}

\maketitle

\section{Summary}
We implemented and optimized matrix multiplications between dense and block-sparse matrices on CUDA. We leveraged TVM, a deep learning compiler, to explore the schedule space of the operation and generate efficient CUDA code. With the automatic parameter tuning in TVM, our cross-thread reduction based implementation achieved competitive or better performance compared with other state-of-the-art frameworks. 

Our code is available here: \href{https://github.com/ceruleangu/Block-Sparse-Benchmark}{https://github.com/ceruleangu/Block-Sparse-Benchmark}

\section{Introduction}

Contemporary deep learning researches require efficient GPU kernels to perform intense computations such as model training and inferences. General matrix multiplication (GEMM) plays an essential role in the computation of deep neural networks because both convolution operations and fully connected operations can both be represented through GEMM. To accelerate neural network computation, the sparsity of weights has been widely utilized. For example, fully connected operations do not scale well because the weights learned for a layer are dense matrices whose sizes depend on not only the size of the input layer but also that of the output layer, while sparse fully connected operations have computational complexity only proportional to the number of non-zero elements. However, sparse operations for arbitrary sparsity cannot be efficiently implemented on current GPUs because the highly parallelized computations of GPU cannot align with the sparsity patterns. In recent literature, block-sparse operations have gradually come to our sights. They have been successfully applied to different domains such as computer vision \citep{DBLP:journals/corr/XieGDTH16, DBLP:journals/corr/ZhangZLS17} and natural language process \citep{DBLP:journals/corr/SakSB14, DBLP:journals/corr/KuchaievG17}.

\begin{figure}
    \centering
    \includegraphics[width=1.1\textwidth]{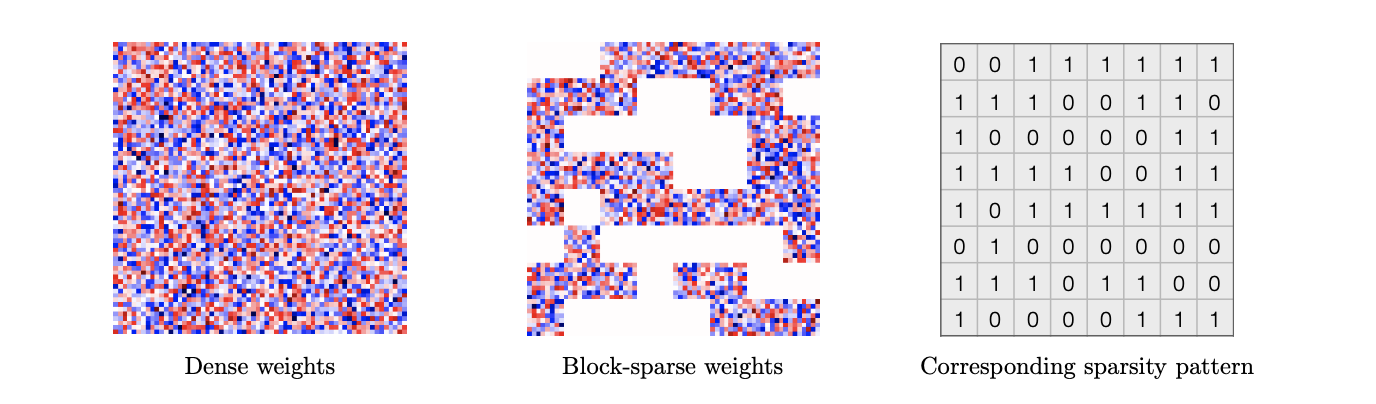}
    \caption{Matrix with dense entries (left), matrix with block-sparse entries (middle), sparsity pattern of the block-sparse matrix in the middle (right).}
    \label{fig:sparse}
\end{figure}

Figure \ref{fig:sparse} \citep{Gray2017GPUKF} shows an example of block-sparse matrix, where the empty blocks in the middle figure indicate all-zero-element blocks. Thus we can represent the sparsity pattern in a matrix as shown in the rightmost figure in Figure \ref{fig:sparse}. In this project, we will be specifically implementing sparse-dense matrix multiplications, in which case the inputs include a dense matrix and a block-sparse matrix, and the output is a dense matrix. This setting has been commonly adopted in the operations of deep neural networks. 

\section{Background}\label{background}
The key data structure for this task is the block-sparse matrix. A block-sparse matrix is stored in block compressed row (BSR) format, which is similar to compressed sparse row (CSR). BSR has three components: 

\begin{enumerate}
    \item \lstinline{block_data}: a three-dimensional array with shape [number of non-zero block, block row number, block column number]
    \item \lstinline{block_indices}: an array of column indices of each block
    \item \lstinline{index_pointer}: an array of indices of the elements in the \lstinline{block_indices} array that are the first blocks of each matrix row 
\end{enumerate}

The key operation we are optimizing is sparse-dense matrix multiplications $Y = XW^\intercal$, where $X$ is a dense matrix with shape $(m, k)$, $W$ is a transposed block-sparse matrix with shape $(n,k)$ in BSR format and $Y$ is the output matrix with shape $(m, n)$. Note that $W$ is transposed from the original matrix to improve the memory locality during computation. Similar to GEMM for dense matrices, we can parallelize the block-sparse matrix multiplications by computing each element in the output matrix concurrently. However, unlike GEMM for dense matrices, the block-sparse matrix in BSR format involves indirect indexing at the runtime: We need to read \lstinline{index_pointer} and \lstinline{block_indices} parts of $W$ and then find the corresponding part in the dense matrix $X$. This makes memory-access patterns unpredictable and thus it is difficult to apply memory optimization methods such as utilizing shared memory as cache.


We implemented the sparse-dense matrix multiplication on TVM \citep{222575}, a deep learning compiler. TVM uses its intermediate representation to describe and optimize tensor operations. It provides a set of domain-specific languages (DSL) API such as loop tiling and loop re-order. With TVM, we can implement the operation with the DSL API in Python and then generate efficient CUDA code. This allows us to explore different implementations for this task efficiently. The implementation of a TVM operation has two parts that are defined separately: the computation and the schedule. The computation part is to describe the operation in a tensor expression language in a way similar to Halide \citep{Ragan-Kelley:2013:HLC:2491956.2462176}. Listing \ref{exampletvm} shows an example of the tensor expression in TVM. The schedule part is to use the provided schedule primitives to map from a tensor expression to low-level code while preserving the logical equivalence of the program. 

\begin{center}
    \begin{lstlisting}[caption={Example of tensor expression in TVM}, captionpos=b, label={exampletvm}]
        m, n, h = tvm.var('m'), tvm.var('n'), tvm.var('h')
        A = tvm.placeholder((m, h)), name='A')
        B = tvm.placeholder((n, h)), name='B')
        k = tvm.reduce_axis((0, h)), name='k')
        C = tvm.compute((m,n), lambda y,x: tvm.sum(A[y,k]*B[x,k],axis=k))
    \end{lstlisting}

\end{center}

\section{Approaches}\label{sec:approach}

As introduced in Section \ref{background}, we first implemented the block-sparse matrix multiplications with TVM tensor expressions DSL to define the computational semantics. And then, we used the schedule primitives of TVM to explore the schedule space of the block-sparse operations on CUDA. After that, we integrated AutoTVM, the machine-learning-based optimizer of TVM, to automatically tune the parameters of our schedules, such as the tiling size of the loops. We implemented and tested on AWS one g4dn.xlarge instance which contains one NVIDIA T4 GPU. We denote the operation we are optimizing as the following:

Let data $X \in \mathbb{R}^{m\times k}$, block $B \in \mathbb{R}^{b_r\times b_c}$, and transposed weights $W_B\in \mathbb{R}^{n\times k}$ where $b_r | n$, $b_c|k$. We want to optimize the time of computing

\begin{equation*}
    XW^\intercal_B = Y \in \mathbb{R}^{m\times n}
\end{equation*}
Note that in reality, $W_B$ is stored in BSR format which consists of  \lstinline{block_data}, \lstinline{block_indices} and \lstinline{index_pointer} introduced in Section \ref{background}

\subsection{Per-element parallelization}\label{baseline}\label{sec:pep}
Our first approach is to parallelize the computation by using one thread for each output element $Y_{ij}$. Specifically, we use a total of one thread block that contains $m\times n$ threads. However, we find that when $m\times n$ is large, a runtime error occurs because the thread block explodes with too many threads in it. We solved the problem by using $m\times n$ thread blocks, each of which contains one thread. The result shows that the runtime increases as sparsity decreases. We think this is because as sparsity decreases each thread needs more data reads from $W_B$ and all these global memory accessing bring a lot of overhead. 

\subsection{Per-tile parallelization}\label{shared}\label{sec:ptp}
Since consecutive $Y_{ij}$ share the same part of $X$ and $W_B$ during computation, similar to the optimization for GEMM for dense matrices, we divide the $m\times n$ output elements in $Y$ into multiple tiles of the same size and assign each tile to a different thread.
We think that assigning several consecutive $Y_{ij}$ to the same thread can reduce some overhead from global memory access. Even though the computations of the $Y_{ij}$ in one tile are no longer concurrent, it may benefit from the locality of accessing $X$ and $W_B$ from the global memory.
In the experiment, we tried several sizes of tiles. However, the result gets worse compared with the first approach. We think it is because the overhead from sequential computation exceeds that from the global memory access.

\subsection{Utilizing shared memory}
To reduce global memory access and preserve locality without losing concurrency in computations, we choose caching as our third approach. We divide the $m\times n$ output elements in $Y$ into multiple tiles of the same size, like in Section \ref{shared}. But instead of assigning each tile to a different thread, we assign the computation for each tile to a different thread block. For each element in one tile, we use a thread for the computation.
In this way, threads within a thread block have locality in accessing $X$ and $W_B$. Therefore, we use shared memory as the cache for the part of $X$ required by all $Y_{ij}$ in the same tile, as well as the shared $W_B$, block index, and pointers.

Specifically, before computing the output, threads will copy the needed part of $X$, \lstinline{block_data}, and \lstinline{block_indices} into the shared memory through cooperative fetching, and then perform \lstinline{__syncthreads}. 
However, note that in sparse-dense matrix multiplication, the needed parts of $X$ and \lstinline{block_data} for the computation is dependent on \lstinline{block_indices}, which involves checking the indices at the runtime and then choose the corresponding parts of $X$ and \lstinline{block_data} to be cached. This indirect indexing introduces additional overhead compared with dense matrix multiplication. 

In our implementation of this caching approach with TVM, we met another issue: TVM requires all caching to be decided at compile time. Since $W_B$ is a runtime value and the compiler does not know which parts of $W_B$ are dense, the entire rows of $X$ are copied to shared memory even though we do not need the parts of $X$ that corresponds to the sparse parts of $W_B$ for the computations. It is a TVM limitation that runtime caching is not viable. Therefore, we are not able to implement this approach.

\subsection{Concurrent reduction}\label{sec:parallel_reduction}

Based on previous analysis, we decide to discard using shared memory and to limit memory access to only the parts of $X$ that corresponds to the dense parts of $W_B$ while preserving concurrency in computation. In Section \ref{baseline}, we use one thread to compute one $Y_{ij}$ in one thread block. In this section, we parallelize the reduction of computing one $Y_{ij}$. Specifically, we use $m\times n$ thread blocks, each of which contains multiple threads to perform the reduction concurrently for a single $Y_{ij}$. In sparse-dense matrix multiplication, the reduction for $Y_{ij}$ is the sum of the products between each non-sparse element in the $j$-th row of $W_B$ and its corresponding element in the $i$-th row of $X$. The details of reduction across different threads are provided in Section \ref{reduction}. We explore two ways of parallelizing reduction in Section \ref{over} and Section \ref{within}. 

\subsubsection{Parallelize over blocks}\label{over}\label{sec:prob}

\begin{figure}[h]
    \centering
    \begin{subfigure}[b]{0.45\textwidth}
        \centering
        \includegraphics[width=\textwidth]{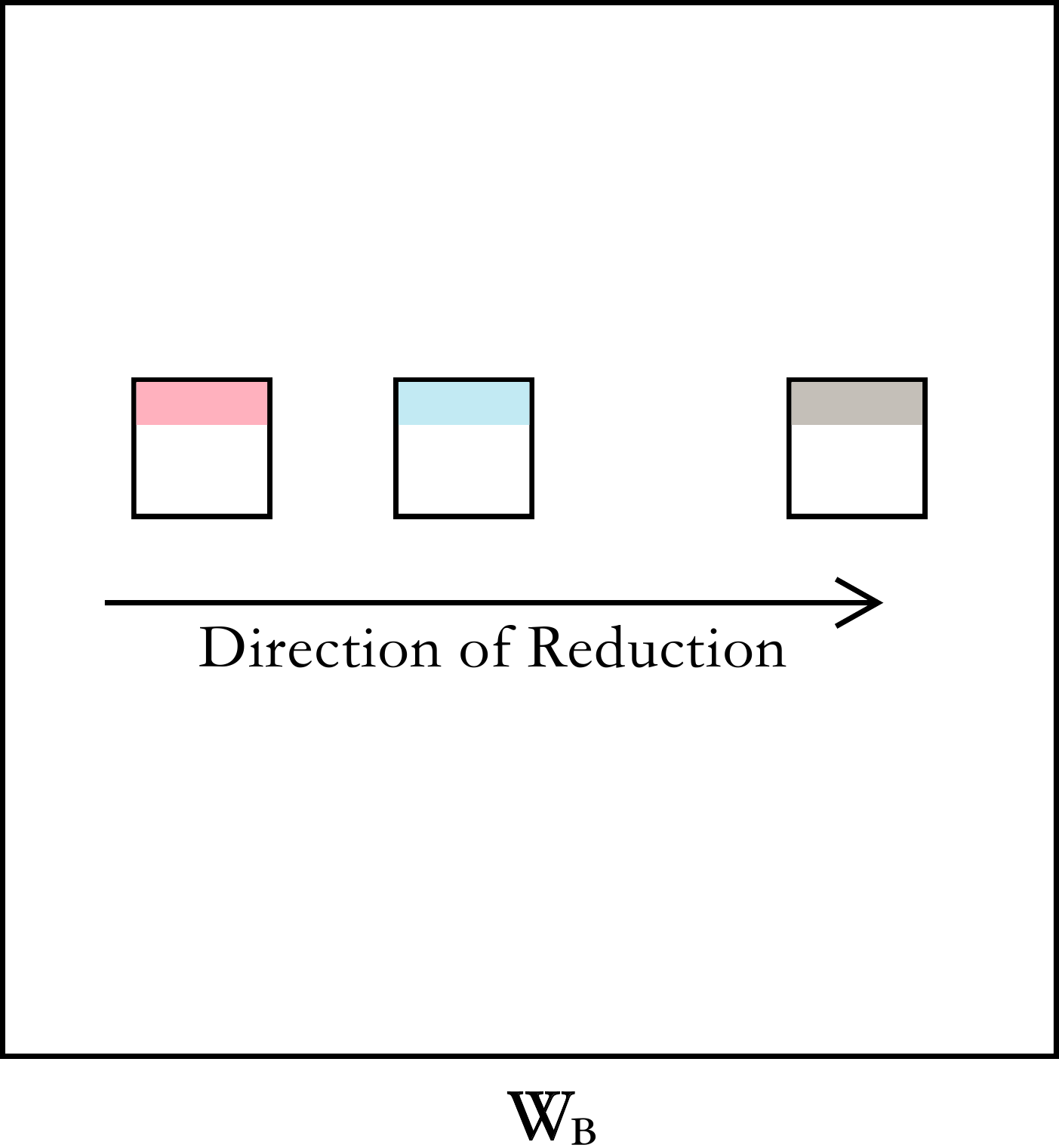}
        \caption{Parallelize over blocks}
        \label{fig:reduction1}
    \end{subfigure}
    \hfill
    \begin{subfigure}[b]{0.45\textwidth}
        \centering
        \includegraphics[width=\textwidth]{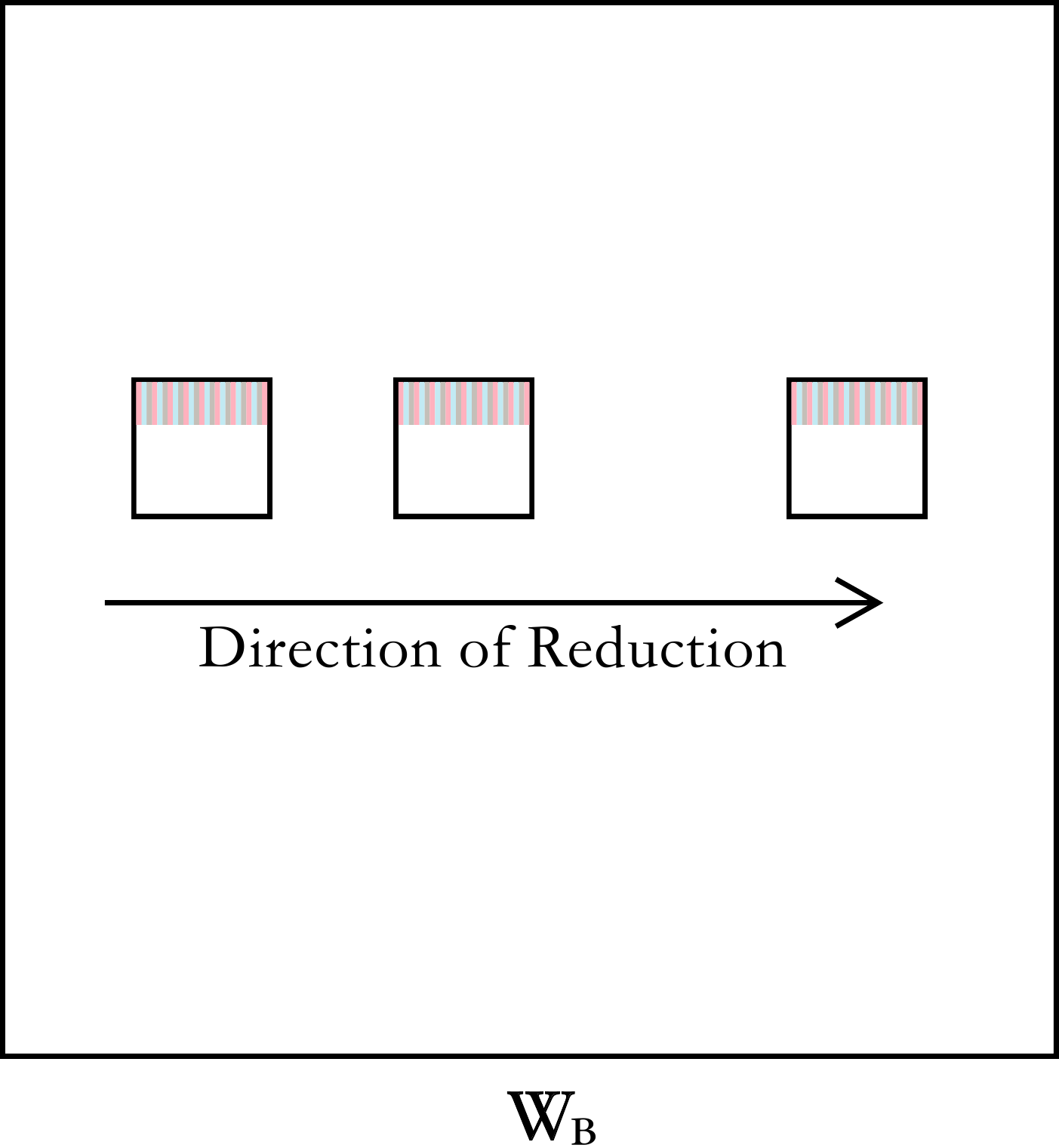}
        \caption{Parallelize within blocks}
        \label{fig:reduction2}
    \end{subfigure}
    \caption{Example of parallel reduction}
\end{figure}

In this approach, we parallelize reduction for $Y_{ij}$ in the following method: for each non-sparse block $B$, we assign a thread to perform reduction within $B$.
As shown in Figure \ref{fig:reduction1}, each color in the figure represents a different thread.
Ideally, we want to set the number of threads in each thread block to be the number of non-sparse $B$ in each row of $W_B$, because we do not need the sparse parts to compute $Y_{ij}$. However, since $W_B$ is a runtime value and the number of non-sparse $B$ in each row of $W_B$ varies, we have to set the number of threads in a thread block to be the total number of $B$ in that row, which is $k/b_c$ (if $k/b_c$ is large, one thread may take charge of multiple $B$). Note that this may exceed the actual number of the threads needed for the computation. Therefore, we use an if-statement to decide whether a thread should be idle or not depending on whether the $B$ assigned to that thread is sparse or not. But this incurs divergence of control flow. As expected, the results do not improve.

\subsubsection{Parallelize within blocks}\label{within}\label{sec:prwb}
In Section \ref{sec:prob}, different threads perform reductions on different $B$ in each row of $W_B$. In this approach, different threads perform reduction within one $B$ and across all dense $B$ in each row of $W_B$. Specifically, we use a fixed number of threads $t$ in each thread block to sum up a portion of $Y_{ij}$ in each dense $B$ of that row in $W_B$. The final $Y_{ij}$ is computed by summing up all the results. 
As shown in Figure \ref{fig:reduction2}, each color in the figure represents a different thread. During the reduction, different threads are accessing consecutive elements within one $B$ from the global memory, which leads to global memory coalescing and improves memory efficiency.

We see improvements in some cases compared with the results in Section \ref{baseline}. We believe that this approach very likely leads to the right direction for optimization. Since it is possible that the number threads per thread block $t$ we choose is not optimal, we apply automatic parameter tuning with TVM in Section \ref{autotvm}.

\subsubsection{Aggregate Partial Reduction results from different threads}\label{reduction}
This section elaborates on the cross-thread reduction. To produce the output $Y_{ij}$, we sum up the partial reduction results from different threads and store those results in the first thread of each thread block. 

\begin{figure}[h]
    \centering
    \includegraphics[width=0.6\textwidth]{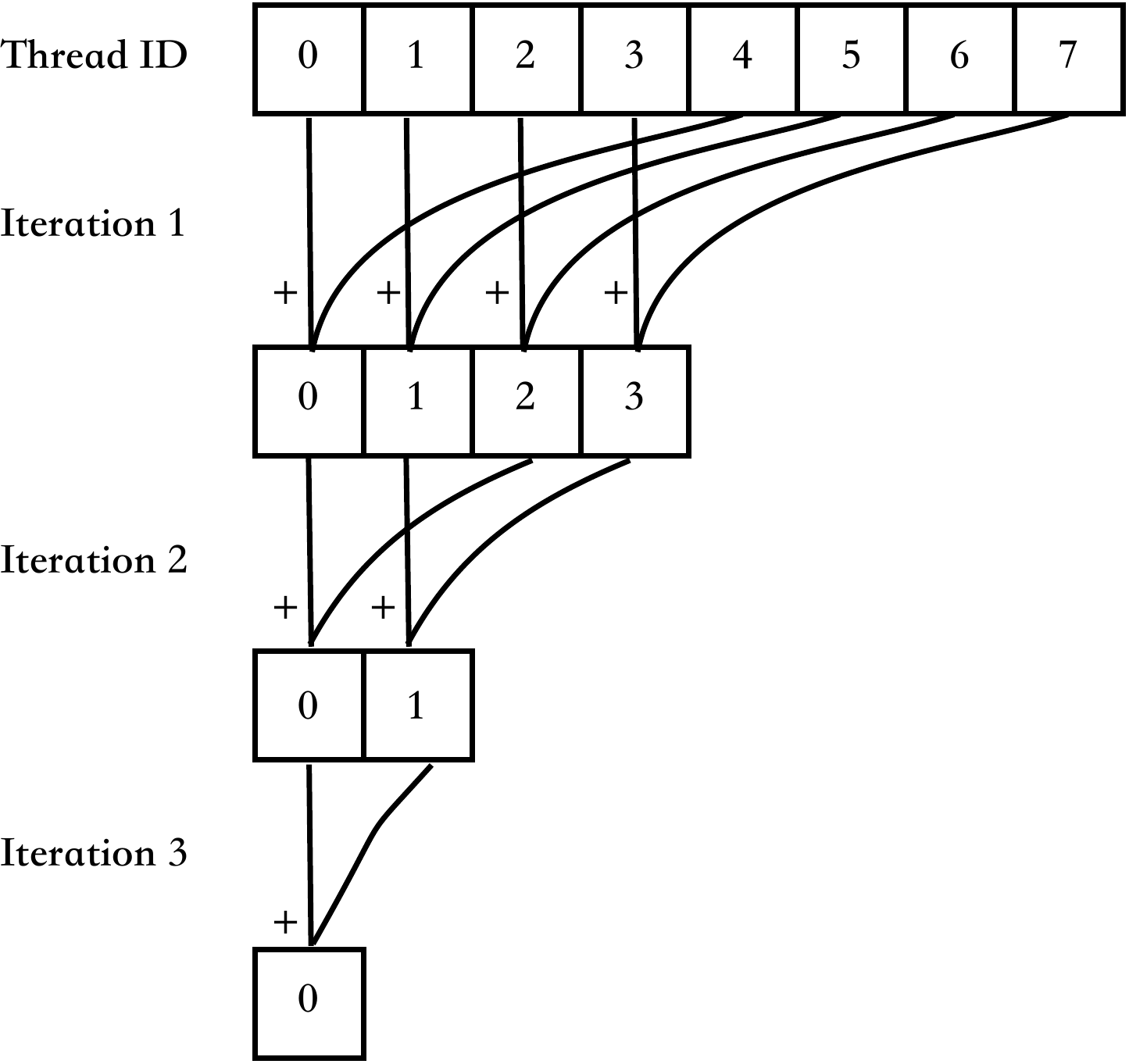}
    \caption{Aggregate results from different threads}
    \label{fig:allreduce}
\end{figure}

We use shared memory to share the results between threads within the same thread block and perform several iterations of reduction as shown in Figure \ref{fig:allreduce}.

When the number of threads in a thread block is 32, we perform additional optimization to eliminate the usage of shared memory. We use \lstinline{__shfl_down} instruction to gather the value from the other threads in the same warp into the first thread.

\subsection{Automatic parameter tuning with TVM}
\label{autotvm}
To search for the optimal number of threads $t$ mentioned in Section \ref{sec:prwb}, we apply TVM to perform automatic parameter tuning for different shapes of input $X$ and $W_B$ and then generate specialized CUDA kernels. We define a list of candidate values of $t$ to be all possible factors of $k$.
TVM will choose certain values from those candidates and generate CUDA code to profile on GPU. Then TVM trains a decision-tree based cost model to predict the performance without running on the actual devices and therefore, the search is very efficient. We perform a maximum of 200 iterations of search which normally finishes within two or three minutes.

\section{Results}
We test our approaches introduced in Section \ref{sec:approach} with different input sizes ($m, k, n$), block sizes ($b_r = b_c$), and sparsity. The test cases are selected from popular deep learning models of computer vision and natural language processing. The testing environment is an NVIDIA T4 GPU on CUDA 10. As shown in Table \ref{tab:bench1}, PEP (Section \ref{sec:pep}) achieves good performance in most cases. Parallel-reduction based methods, PROB and PRWB, (Section \ref{sec:parallel_reduction}) outperform PEP in most cases where $(m, k, n)$ is small. But as $(m, k, n)$ increases, the running time of parallel-reduction based methods drastically increases. Since the possible options for the schedule parameters (such as the number of threads in a thread block) increase as $(m, k, n)$ increases, it is very likely that the current parameters we pick for this benchmark are not optimized. Since PROB suffers divergence of control-flow (Section \ref{sec:prob}), we decide to choose PRWB for AutoTuning (Section \ref{autotvm}). 

After AutoTuning, we achieve the best results among all our approaches. We compare our best results with two state-of-the-art (SOTA) frameworks in Table \ref{tab:bench2}. \citet{Gray2017GPUKF} is a highly-optimized library for block-sparse matrix operations based on manually-tuned micro-kernels, which allows fine-grained control of the instruction orders and achieves competitive performance. cuSparse is the vendor-provided library for sparse and block-sparse matrices that utilizes TensorCore. As shown in Table \ref{tab:bench2}, our approach outperforms SOTA in half of the cases. When ($m, k, n$) is small, the AutoTuning on PRWB generates the best results and is even better than TensorCore-based results. As ($m, k, n$) increases, our results are worse than others. We suggest that the search space of the schedule parameter can be improved so that the AutoTuning might yield better results. We also noticed that our results tend to outperform SOTA when sparsity is extremely high. We suggest that cuSparse is unable to utilize TensorCore very well in those scenarios.


\begin{table}[htbp]
    \centering
    \resizebox{0.85\textwidth}{!}{%
    \begin{tabular}{c|c|c|c|c|c|c}
    \toprule
    $(m, k, n)$ & $B$ size & Sparsity & PEP & PTP & PROB & PRWB \\ \midrule
    \multirow{9}{*}{$(1, 128, 768)$} & \multirow{3}{*}{8} & 0.8 & 0.019 & 0.14 & 0.013 & 0.0091  \\ 
     &  & 0.85 & 0.015 & 0.11 & 0.012 & 0.0081\\ 
     &  & 0.95 & 0.010 & 0.043& 0.011&  0.0071 \\  \cmidrule{2-7}
     & \multirow{3}{*}{16} & 0.8 & 0.025 & 0.18 &0.013 & 0.0084\\ 
     &  & 0.85 & 0.021 & 0.12 &0.013 & 0.0071\\ 
     &  & 0.95 & 0.010 & 0.075 & 0.010 &  0.0076\\ \cmidrule{2-7}
     & \multirow{3}{*}{32} & 0.8 & 0.043 &0.20 & 0.012 & 0.0070 \\ 
     &  & 0.85 & 0.032 &0.17 & 0.012 &  0.0071\\ 
     &  & 0.95 & 0.014 & 0.11&0.010 & 0.0070 \\ \midrule
    \multirow{9}{*}{$(8, 128, 768)$} & \multirow{3}{*}{8} & 0.8 & 0.018 & 0.14& 0.033&  0.030\\ 
     &  & 0.85 & 0.015 & 0.099 & 0.029 &  0.028 \\  
     &  & 0.95 & 0.010 & 0.047 & 0.020 &  0.025 \\ \cmidrule{2-7}
     & \multirow{3}{*}{16} & 0.8 & 0.025 & 0.15 & 0.021 & 0.029\\ 
    &  & 0.85 & 0.020 &0.13 & 0.018 & 0.027\\ 
     &  & 0.95 & 0.011 & 0.050&0.012 & 0.025\\ \cmidrule{2-7}
     & \multirow{3}{*}{32} & 0.8 & 0.043 & 0.28 & 0.015 & 0.030\\ 
     &  & 0.85 & 0.033 &0.21  &0.013 & 0.029\\ 
     &  & 0.95 & 0.015 & 0.077 & 0.012 & 0.025\\ \midrule
    \multirow{9}{*}{$(1, 1024, 1024)$} & \multirow{3}{*}{8}& 0.8 & 0.14 & 0.57 & 0.046 & 0.016 \\
     & & 0.85 & 0.11 & 0.45  & 0.046& 0.014 \\
     & & 0.95 & 0.042 & 0.16  & 0.045&0.010 \\ \cmidrule{2-7}
      & \multirow{3}{*}{16}& 0.8 & 0.22 & 0.64 &0.039 & 0.015\\
     & & 0.85 & 0.16 & 0.48  &0.039 & 0.014\\
     & & 0.95 & 0.058 & 0.19  & 0.038&0.010 \\ \cmidrule{2-7}
      & \multirow{3}{*}{32}& 0.8 & 0.42 & 0.79 & 0.036&0.014 \\
     & & 0.85 & 0.31 & 0.68  & 0.036 &0.013 \\
     & & 0.95 & 0.11 & 0.27  &  0.035& 0.009\\ \midrule
     \multirow{9}{*}{$(8, 1024, 1024)$} & \multirow{3}{*}{8}& 0.8 &0.14 &0.58 &0.23 &0.087 \\
     & & 0.85 & 0.11 &0.44 & 0.20&0.071 \\
     & & 0.95 & 0.041 & 0.16 & 0.14& 0.046\\ \cmidrule{2-7}
      & \multirow{3}{*}{16}& 0.8 & 0.21 & 0.64 & 0.27& 0.079\\
     & & 0.85 & 0.16 &0.49  & 0.26&0.065 \\
     & & 0.95 & 0.058 & 0.20 &0.26 &0.045 \\ \cmidrule{2-7}
      & \multirow{3}{*}{32}& 0.8& 0.42 & 0.79& 0.24 & 0.070\\
     & &0.85 & 0.32 &0.62 & 0.24 &0.059 \\
     & & 0.95& 0.11& 0.28 & 0.23& 0.040 \\ \bottomrule
    \end{tabular}
    }
    \caption{Benchmark of matrix multiplications between dense matrix of size $m\times k$ and block-sparse matrix of size $k\times n$ with different block size and sparsity. ``PEP": Per Element Parallization (Section \ref{sec:pep}). ``PTP": Per Tile Parallization (Section \ref{sec:ptp}). ``PROB": Parallel Reduction Over Blocks (Section \ref{sec:prob}). ``PRWB": Parallel Reduction With Blocks (Section \ref{sec:prwb}). Results are in milliseconds. }
    \label{tab:bench1}
\end{table}

\begin{table}[htbp]
    \centering
    \begin{tabular}{c|c|c|c|c|c}
    \toprule
    $(m, k, n)$ & $B$ size & Sparsity & PRWB+AT(Ours)
    & \citet{Gray2017GPUKF} & cuSparse  \\ \midrule
    \multirow{9}{*}{$(1, 128, 768)$} & \multirow{3}{*}{8} & 0.8 & \textbf{0.0051} & 0.015 & 0.0065\\ 
     &  & 0.85 & \textbf{0.0036} & 0.010 & 0.0060\\ 
     &  & 0.95 & \textbf{0.0036} & 0.008 & 0.0060\\  \cmidrule{2-6}
     & \multirow{3}{*}{16} & 0.8 & \textbf{0.0040}& 0.015 & 0.0061\\ 
     &  & 0.85 & \textbf{0.0037} & 0.013 & 0.0058\\ 
     &  & 0.95 & \textbf{0.0037}& 0.010 & 0.0059\\ \cmidrule{2-6}
     & \multirow{3}{*}{32} & 0.8 & \textbf{0.0036} & - & 0.011\\ 
     &  & 0.85 & \textbf{0.0036} & - & 0.011\\ 
     &  & 0.95 &\textbf{0.0037} & - & 0.011\\ \midrule
    \multirow{9}{*}{$(8, 128, 768)$} & \multirow{3}{*}{8} & 0.8 &0.010 & 0.014 & \textbf{0.0078}\\ 
     &  & 0.85 &0.0085 & 0.011 & \textbf{0.0078}\\  
     &  & 0.95 &\textbf{0.0051} & 0.011 & 0.0059\\ \cmidrule{2-6}
     & \multirow{3}{*}{16} & 0.8 &0.0070 & 0.016 & \textbf{0.0063}\\ 
    &  & 0.85 & 0.0083& 0.014 & \textbf{0.0060}\\ 
     &  & 0.95 & \textbf{0.0047}& 0.010 &0.0060\\ \cmidrule{2-6}
     & \multirow{3}{*}{32} & 0.8 & \textbf{0.010} & - & 0.011\\ 
     &  & 0.85 & \textbf{0.0043} & - & 0.011\\ 
     &  & 0.95 & \textbf{0.0048} & - & 0.011\\ \midrule
    \multirow{9}{*}{$(1, 1024, 1024)$} & \multirow{3}{*}{8}& 0.8 & 0.013 & 0.015 & \textbf{0.0065} \\
     & & 0.85 & 0.011& 0.012 & \textbf{0.0060}\\
     & & 0.95 & \textbf{0.0059} & 0.009 & \textbf{0.0059}  \\ \cmidrule{2-6}
      & \multirow{3}{*}{16}& 0.8 & 0.013 & 0.018&\textbf{0.0062} \\
     & & 0.85 & 0.011& 0.012& \textbf{0.0060}\\
     & & 0.95 & \textbf{0.0058}& 0.008& 0.0060\\ \cmidrule{2-6}
      & \multirow{3}{*}{32}& 0.8 &0.012 & -&\textbf{0.011} \\
     & & 0.85 & \textbf{0.0047} & -&0.011 \\
     & & 0.95 & \textbf{0.0042}& -&0.011 \\ \midrule
     \multirow{9}{*}{$(8, 1024, 1024)$} & \multirow{3}{*}{8}& 0.8 &0.078 & 0.015& \textbf{0.0078}\\
     & & 0.85 & 0.061& 0.013& \textbf{0.0061}\\
     & & 0.95 & 0.026& 0.008&\textbf{0.0060}, \\ \cmidrule{2-6}
      & \multirow{3}{*}{16}& 0.8 &0.074 & 0.019 &\textbf{0.0079} \\
     & & 0.85 & 0.058& 0.014 &\textbf{0.0079} \\
     & & 0.95 & 0.025& 0.009 & \textbf{0.0062}\\ \cmidrule{2-6}
      & \multirow{3}{*}{32}& 0.8& 0.018& - & \textbf{0.011}\\
     & &0.85 & 0.017 & - & \textbf{0.011} \\
     & & 0.95&0.014 & - &  \textbf{0.011}\\ \bottomrule
    \end{tabular}
    \caption{Benchmark of matrix multiplications between dense matrix of size $m\times k$ and block-sparse matrix of size $k\times n$ with different block size and sparsity. ``PRWB+AT":Parallel Reduction Within Blocks + AutoTuning (Section \ref{sec:prwb}, \ref{autotvm}). Results are in milliseconds. }
    \label{tab:bench2}
\end{table}

\section{Conclusion}
We implemented multiplication between dense matrix and block-sparse matrix in BSR format on CUDA with TVM. We explored different schedules using TVM and evaluated their performances. With automatic parameter tuning with TVM, we achieved competitive performances in the benchmark.

\section{Discussion}
In this project, we tried several implementations of sparse-dense matrix multiplication. Since the operation is memory intensive, the performance we achieved is still far below the limit of the computation performance. Even though the parallel reduction based implementations achieved the best performances, the number of threads needed is highly dependent on the sparsity pattern. If a single row in $W_B$ has many non-sparse blocks while some other rows are highly sparse, such imbalance will incur divergence of control flow since the number of iterations for the reduction differs. As a result, the performance of parallel-reduction based implementation is significantly worsened, and the simple parallelization approach in Section \ref{baseline} will suffice. Hence, we believe that there is no single implementation that has the best performance in all cases. To achieve good performance, we may need to combine different approaches for different cases.

Additionally, we observe that the absolute performance of both our methods and SOTA are very low. The peek FLOPS is about 400GFLOPS, which is far below the theoretical performance of NVIDIA T4 GPU (8.1TFLOPS). We think the performance of the sparse-dense matrix multiplication is mainly limited by memory access, and therefore it is difficult to gain further performance improvement.

\bibliographystyle{unsrtnat}
\bibliography{bib.bib}
\end{document}